\documentclass[twocolumn,preprintnumbers,superscriptaddress,nofootinbib,aps,prd,floatfix]{revtex4}

\usepackage{footmisc}
\usepackage{amsmath}
\usepackage{wasysym} 
\usepackage{graphicx}
\usepackage{color}
\usepackage{comment}

\newcommand{\be}{\begin{eqnarray}}
\newcommand{\ee}{\end{eqnarray}}

\newcommand{\bee}{\begin{eqnarray}}
\newcommand{\eee}{\end{eqnarray}}
\newcommand{\beeq}{\begin{equation}}
\newcommand{\eeeq}{\end{equation}}
\newcommand{\gev}{{\text{GeV}}}

\begin{document}

\title{Gluon-initiated associated production boosts Higgs physics}

\begin{abstract}
  Analyses of boosted Higgs bosons from associated production comprise
  some of the main search channels for the Higgs boson at
  the LHC.  The gluon-initiated $gg\to hZ$ subprocess has largely been
  ignored in phenomenological analyses of boosted associated
  production although this contribution is sizable as the $p_T$
  spectrum for this process is maximised in the boosted regime due to
  the top quark loop threshold. In this paper, we discuss this
   contribution to boosted $pp\to hZ$ analyses in detail.
  We find there are previously overlooked modifications of Standard
  Model Higgs rates at the LHC which depend on the $p_T$ cuts applied and can be significant.
  There are also important consequences for physics beyond the Standard
  Model as the $gg\to hZ$ process introduces significant dependence on
  the magnitude and sign of the Higgs-top quark coupling $c_t$, which
  is overlooked if it is assumed that associated production depends
  only on the Higgs-$Z$ boson coupling as $c_V^2$.  This new
  dependence on $c_t$ impacts interpretations of Higgs rates in the
  contexts of Supersymmetry, Two Higgs Doublet Models, and general
  scenarios with modified couplings.  We suggest that
  these effects be included in current and future LHC boosted Higgs
  analyses.
\end{abstract}

\author{Christoph Englert} \email{christoph.englert@glasgow.ac.uk}
\affiliation{SUPA, School of Physics and Astronomy, University of
  Glasgow,\\Glasgow, G12 8QQ, UK \\[0.1cm]}
\author{Matthew McCullough} \email{mccull@mit.edu} \affiliation{Center
  for Theoretical Physics, Massachusetts Institute
  of Technology,\\
  Cambridge, MA 02139, USA \\[0.1cm]}
\author{Michael Spannowsky} \email{michael.spannowsky@durham.ac.uk}
\affiliation{Institute for Particle Physics Phenomenology, Department
  of Physics,\\Durham University, DH1 3LE, UK}

\preprint{DCPT/13/158}
\preprint{IPPP/13/79}
\preprint{MIT-CTP 4506}
\maketitle

\section{Introduction}
\label{sec:intro}
The discovery of a boson at the LHC
\cite{Chatrchyan:2012ufa,Aad:2012tfa} largely consistent with the
particle resulting from the Standard Model (SM) Higgs mechanism
\cite{orig} marks the beginning of a new era of particle physics. For
the first time we are provided with an opportunity to gain a better
understanding of the electroweak scale through precise analyses and
measurements of this newly discovered state. Crucial to the Higgs
agenda is the precise measurement of the Higgs boson couplings to SM
fields. The observed mass $m_h\simeq 125~\gev$ provides us with the
fortunate circumstance that all dominant fermionic and bosonic Higgs
decay channels are accessible at the LHC, and it is possible to probe
the nature of the Higgs boson at $\lesssim 10\%$ precision at high
luminosity~\cite{Plehn:2012iz}.

The measurement of the Higgs-bottom quark coupling is enabled by
exploiting boosted final states in conjunction with recently-developed
subjet technology~\cite{Butterworth:2008iy} in associated production
$pp\to hZ$. The latter is dominated by quark-initiated subprocesses,
but there is also a large gluon-initiated contribution, $gg\to
hZ$~\cite{ggasshz,kniehl}, which has typically not been included in detail in
the corresponding analyses.\footnote{The $gg\to hZ$ contribution can
  easily be missed by adopting an ${\cal{O}}(\alpha_s^2)$ $K$-factor
  normalisation of a matched $q\overline q\to hZ$ sample. This does
  not reflect the differences in the differential characteristics of
  the $gg$ and $q\overline q$ contributions, particularly with regard
  to the $p_T$ spectrum.}  The na\"{\i}ve cross section suppression of
the gluon-initiated subprocesses compared to the quark-initiated
processes by roughly an order of magnitude \cite{asshz} is not
only compensated in part by much larger QCD perturbative corrections
which enhance the role of the gluon-initiated
component~\cite{ggasshzqcd}, but the top quark loop also induces a
scale when absorptive parts of the scattering amplitude open up for
invariant masses of the $hZ$ system $m_{hZ}\gtrsim 2m_t$. It is
straightforward to see that this phase-space region is characterised
by boosted kinematics $p_{T,h}\gtrsim 150~\gev$. Hence, there is major
sensitivity to $gg\to hZ$ in boosted analyses as these effects combine
to lift the na\"{\i}ve cross section suppression of $gg\to hZ$.
Because the gluon-initiated subprocess provides a non-negligible
contribution to the boosted $pp\to hZ$ rate at the LHC additional sensitivity to new physics is introduced with this process \cite{robert}, and there is a
significant impact on future Higgs coupling extractions at high LHC
luminosities through the introduction of significant dependence on the
magnitude and sign of the Higgs-top quark coupling.  Furthermore, the
gluon-initiated component is absent for $pp\to hW$.  Hence, although
the Higgs couplings to $Z$ and $W$ bosons may respect custodial
symmetry to a high degree, i.e.\ $c_Z=c_W=c_V$, the gluon-initiated
contribution means that $pp\to hZ$ and $pp\to hW$ need not respect
this symmetry.\footnote{Throughout we define the Higgs coupling
  factors $c_i$ as the ratio of the Higgs coupling to some SM state to
  the SM value.}  This subtlety is missed if only the $q\overline q\to
hZ, hW$ process is assumed in Higgs coupling fits.  Gluon-initiated
associated production also introduces sensitivity to new coloured
states coupled to the Higgs which enter the $gg\to hZ$ loops.

We will first review $pp\to hZ$ production to make this work
self-contained and subsequently perform a hadron-level analysis of the
final state, including a leading order $gg\to hZ$ sample keeping all
mass and Higgs coupling dependencies.  We discuss the impact of
including the effects of the $gg\to hZ$ process for SM Higgs rates at
the LHC in boosted channels, finding that an accurate estimation of
the cross section in boosted channels requires consideration of the
full $p_T$ distribution of the gluon-initiated contribution, rather
than including this process as a rescaling of the quark-initiated
distribution.\footnote{Even when the event sample is corrected to
  distributions obtained with parton-level Monte Carlos the different
  shower profile of the gluon contribution is not included.} We
also provide fits of the dependence of the associated production cross
section on the Higgs couplings before and after typical selection
cuts, including those relevant to $h\to \overline{b} b$ searches.  We
use this coupling dependence to evaluate the impact of the $gg\to hZ$
process on the extraction of new physics signatures from Higgs
coupling fits.


\section{Gluon-initiated $hZ$ production in the boosted regime}
\label{sec:hz}
Given the importance of associated Higgs boson production, the
gluon-initiated contribution to $hZ$ production was calculated some time
ago \cite{ggasshz,kniehl}. The QCD corrections to this process, however, have
been made available only recently \cite{ggasshzqcd} in the
$m_t\to\infty,m_b\to 0$ approximation. While the quark-initiated
subprocesses follow a Drell-Yan-type paradigm with a moderate
\hbox{(next-to-)}next-to leading order $K$-factor of $K\simeq 1.2$ the
gluon-initiated contribution receives NLO radiative corrections of
\begin{figure}[!t]
  \centering
  \includegraphics[width=0.48\textwidth]{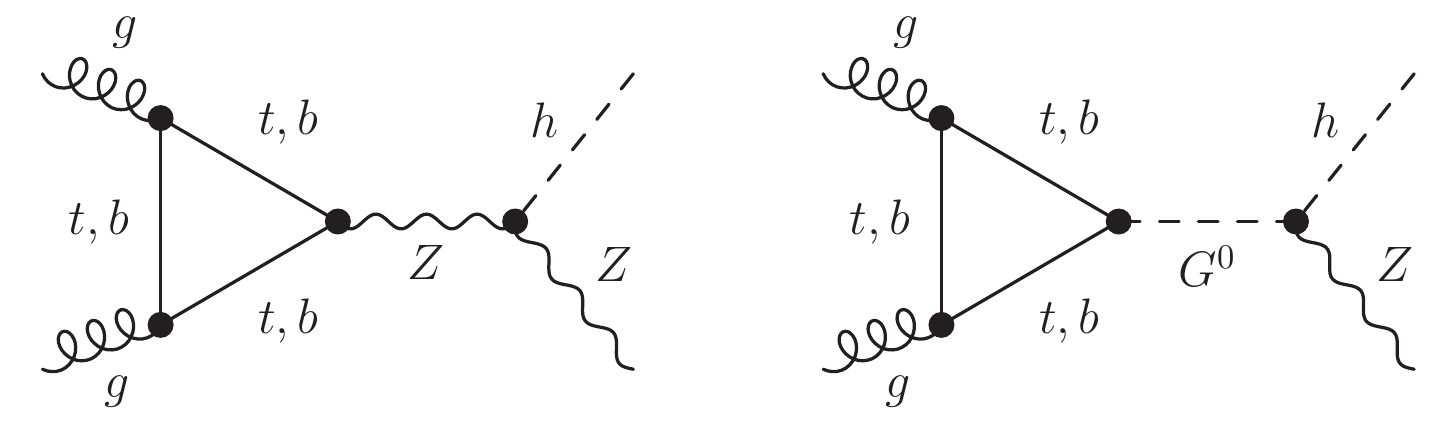}\\[0.3cm]  
  \parbox{0.202\textwidth}{
    \includegraphics[width=0.202\textwidth]{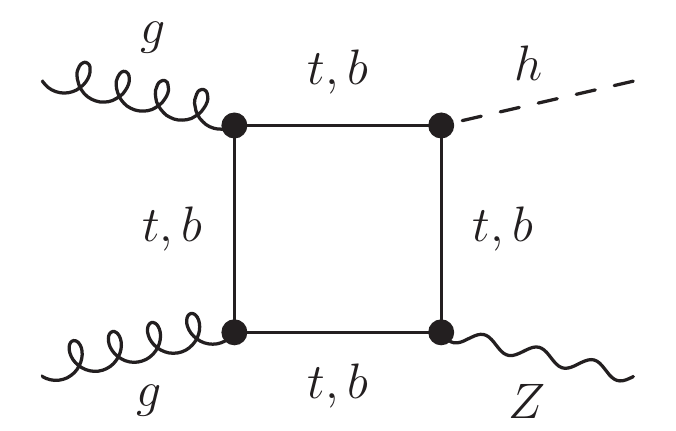}
  } \hfill
  \parbox{0.22\textwidth}{ \caption{\label{fig:feynman} Feynman
      diagram topologies contributing to $gg\to hZ$ at leading order
      in general gauge.}
  }
\end{figure}
$K\simeq 2$, similar to $gg\to h,hh$
production~\cite{Spira:1995rr,Dawson:1998py}, as a consequence of
larger initial state color charge $C_A/C_F=9/4$. We will not delve
into the details of perturbative corrections, but will assume the
total correction factor as reported in \cite{asshz,ggasshzqcd} as flat
in the actual analysis. The characteristic
leading order (LO) features, which are central to the discussion in
this paper will also persist beyond LO.

\begin{figure*}[!t]
  \centering
  \includegraphics[height=0.33\textwidth]{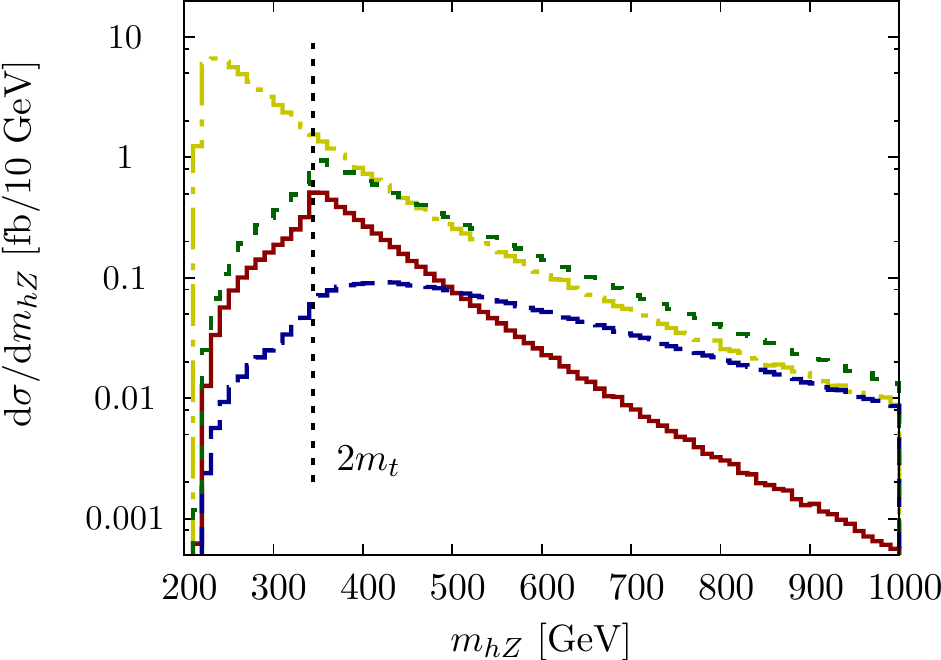}
  \hspace{0.5cm}
  \includegraphics[height=0.33\textwidth]{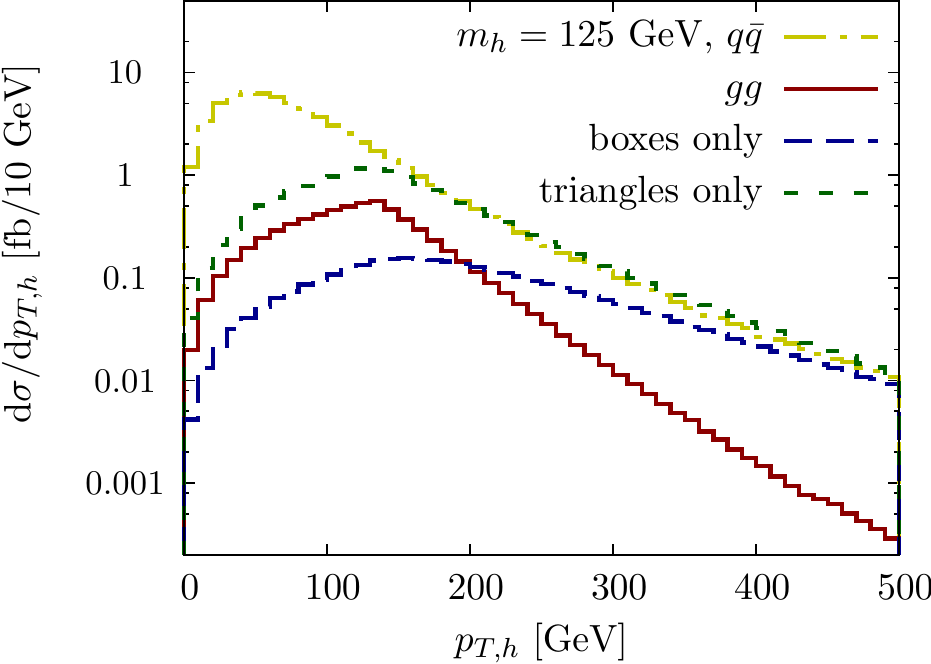}
  \caption{\label{fig:ptsm} Invariant $hZ$ mass $m_{hZ}$ (left) and
    $p_T$ spectra (right) for $pp\to hZ$ production at
    $\sqrt{s}=14$~TeV.  The gluon-initiated and quark-initiated
    contributions are shown for comparison We also plot contributions
    from box and triangle diagrams to demonstrate the cancellation
    between the two in the sum.}
\end{figure*}

\begin{figure}[!t]
  \includegraphics[height=0.33\textwidth]{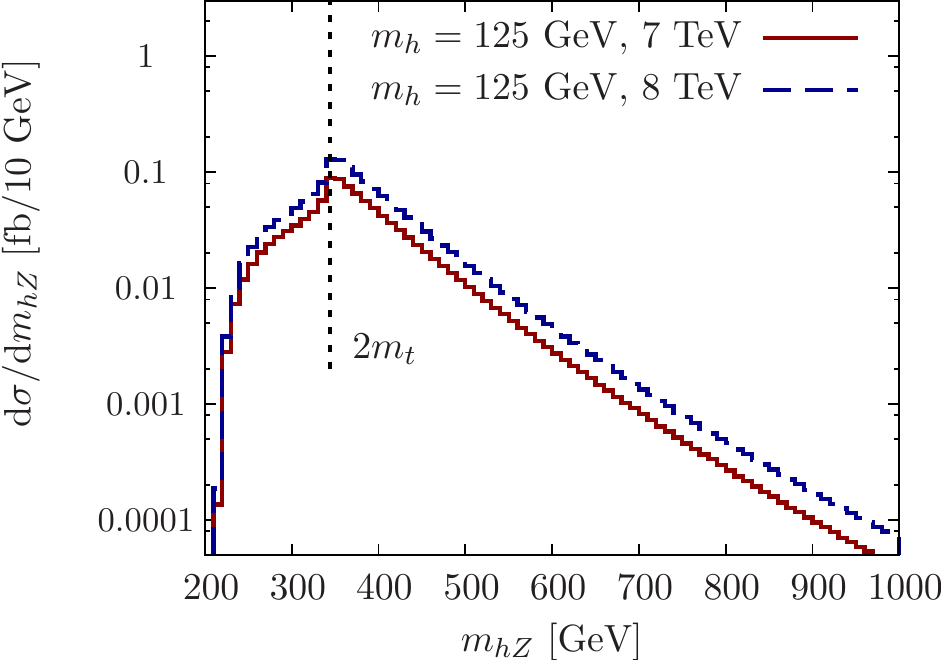}
  \caption{\label{fig:ptsm2} Invariant $hZ$ mass $m_{hZ}$ for the gluon-initiated component of $pp\to
    hZ$ production at $\sqrt{s}=7$~TeV and $\sqrt{s}=8$~TeV for
    comparison with Fig.~\ref{fig:ptsm}.}
\end{figure}

Gluon-initiated associated production is computed from the Feynman
topologies depicted in Fig.~\ref{fig:feynman}. The special role of the
top quark follows from the threshold behaviour of the amplitude which
has a branch cut $s\geq 4m_t^2$, giving rise to an absorptive part of
the amplitude related to other physical process according to the
Cutkosky rules \cite{Cutkosky:1960sp}. This can be seen in
Fig.~\ref{fig:ptsm}, where we compare the different contributions to
$pp\to hZ$ at the LHC for $\sqrt{s}=14$~TeV (see Fig.~\ref{fig:ptsm2}
for 7 TeV and 8 TeV results).\footnote{We have cross-checked these
  results against existing calculations in the literature
  \cite{ggasshz,kniehl,ggasshzqcd,asshz,sherpa,madevent} and find excellent
  agreement.} While this may be considered common knowledge, it is
granted little attention in the estimation of Higgs signal rates and
the coupling extraction effort. This is understandable in the light of
the limited LHC Run~I data which relies on total signal counts and
hence the high $p_{T,h}$ analysis currently has a negligible impact on
Higgs coupling extractions. However, this situation will change
fundamentally with 14 TeV data and the high $p_{T,h}$ analyses will be
central to the Higgs coupling extraction at a high luminosity run
which will crucially rely on exclusive selections and differential
Higgs cross sections.

We calculate the quark-initiated and one loop gluon-initiated
associated production amplitudes using the
{\sc{FeynArts}}/{\sc{FormCalc}}/{\sc{LoopTools}} \cite{hahn}
frameworks.  We use a Monte Carlo calculation based on the
{\sc{Vbfnlo}}~\cite{vbfnlo} framework to generate parton-level events
in the Les Houches standard \cite{lhe} which we pass to
{\sc{Herwig++}} \cite{Bahr:2008pv} for showering and hadronisation.

We apply typical $hZ$ final state selection cuts by requiring exactly
2 oppositely charged same-flavor leptons satisfying $|\eta_l | < 2.5$
and $p_{T,l} > 30$ GeV and with invariant mass in the region $80 <
m(l_1,l_2) < 100$ GeV.  We tag boosted $Z$-boson candidates by
requiring $p_T (l_1+l_2) >200$ GeV.  To reconstruct the Higgs boson in
$h\to \overline{b} b$ we combine jets using the Cambridge-Aachen
algorithm with radius $R=1.2$ and require a boosted Higgs boson
candidate by requiring the jet $p_T$ satisfies $p_{T,j} > 200$ GeV.
At least one fat jet is required with $|\eta_j| < 2.5$ and the
$b$-tagging is applied to this jet.

Jet substructure techniques are implemented as in the BDRS analysis
\cite{Butterworth:2008iy} with a double $b$-tag on the filtered
subjets.  The doubly-tagged reconstructed Higgs jet has to have mass
in the window $115~\text{GeV}<m(\overline{b}b)<135$ GeV.  We impose a
$60\%$ signal tagging efficiency and a $2\%$ fake tagging rate.

\begin{figure}[!b]
  \centering
  \includegraphics[height=0.33\textwidth]{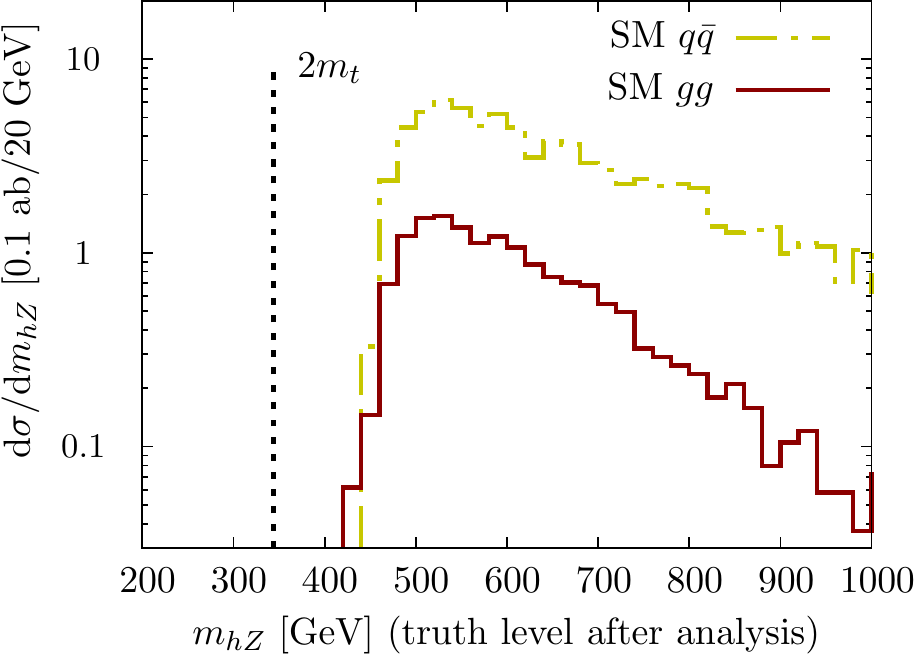}
  \caption{\label{fig:ptsmana} Invariant truth-level $hZ$ mass for
    $pp\to (h\to \overline b b) (Z\to \mu^+\mu^-,e^+e^-)$ production
    in the SM at $\sqrt{s}=14$~TeV. These results are a direct
    reflection of Fig.~\ref{fig:ptsm} after the analysis cuts and the
    reconstruction have been applied. NLO correction factors as
    reported in Refs.~\cite{asshz,ggasshzqcd} have been included to
    reflect the proper signal composition.}
\end{figure}

After the analysis steps described above we find a signal cross
section of $\sigma=0.2~{\rm{fb}}$ which contains the contribution from
the gluon-initiated sample. We also include the relevant $K$ factors
as described earlier. The differential composition before cuts is
shown in Fig.~\ref{fig:ptsm} and after cuts and BDRS analysis is shown
in Fig.~\ref{fig:ptsmana}.

Obviously the boosted selection cuts (which cannot be relaxed unless
the $t\overline t$ backgrounds are suppressed by other means) remove
the $m_t$ threshold behaviour encountered in the $gg$
subprocesses. Nonetheless the contribution is still non-negligible and
the interplay of the box and triangle contributions can be used to
formulate constraints on the involved couplings at large LHC
luminosity.

\section{Implications for SM Rates at the LHC}
\label{sec:SM}
This result has implications for the extraction of SM Higgs rates in
the boosted $pp\to hZ$, $h\to \overline{b}b$ channel.  
Currently rates
are calculated in this channel by applying the selection cuts for
boosted associated production to $p_T$ distributions calculated at NLO
which only include the quark-initiated component.  NNLO corrections
are taken into account by simply applying an overall rescaling to the
distributions with the required $K$-factors, ensuring that the total
associated production cross section matches the NNLO results.
Gluon-initiated $hZ$ is technically NNLO, hence the current methods
overlook the differences in distributions between quark-initiated and
gluon-initiated processes.  These differences are significant, as
demonstrated in Fig.~\ref{fig:ptsm}.  The gluon-initiated $hZ$ distributions at $7$ and $8$ TeV are also shown and exhibit the same qualitative behavior.

Schematically, if we denote the application of typical selection cuts
on an $hZ$ production process at the LHC as $\text{C} \lbrack \sigma
\rbrack$ and the BDRS analysis on the $\overline{b}b$ final state as
$\text{B} \lbrack \sigma \rbrack$, then with current methods employed
at the LHC the boosted associated production cross section after
selection cuts is calculated as
\be
\sigma_{\text{Cuts}} = K^{\text{eff}} \times \text{C} \lbrack 
\sigma_{\overline{q}q} (pp\to hZ) \rbrack  ~~,
\ee
where the subscript $\overline{q}q$ denotes the quark-initiated
process with distributions calculated at NLO.\footnote{Both QCD and EW
  corrections are included at NLO, however, NNLO effects, including
  gluon-initiated associated production, are only applied at the
  inclusive, or total cross-section, level.}  After applying the full
BDRS analysis the resulting cross-section is
\be
\sigma_{\text{BDRS}} = K^{\text{eff}} \times \text{B} \lbrack
\text{C} \lbrack \sigma_{\overline{q}q} (pp \to hZ, h\to
\overline{b} b) \rbrack \rbrack ~~.
\ee
The effective $K$-factor is
calculated from the inclusive cross sections as
\begin{equation}
  \label{eq:keff}
  K_{\text{eff}} = \frac{K_{\overline{q}q}^{\text{NNLO}} \times
  \sigma^{\text{Inc}}_{\overline{q}q} + K_{gg}^{\text{NLO}} \times
  \sigma^\text{Inc}_{gg}}{\sigma^\text{LO,Inc}_{\overline{q}q}} ~~,
\end{equation}
where the superscript ``$\text{Inc}$'' represents the fact that these
quantities are calculated at the inclusive level.  However, because
the differential distributions for the boosted quark-initiated and
gluon-initiated contributions are different they behave differently
under the selection cuts and BDRS analysis, invalidating the approach
sketched above.  To obtain a more accurate result the cuts and BDRS
analysis should be applied to events originating from both production
mechanisms.  Doing this one would calculate
\bee
\widetilde{\sigma}_{\text{Cuts}}  = & K_{\overline{q}q}^{\text{NNLO}}
\times \text{C} 
\lbrack \sigma_{\overline{q}q} (pp\to hZ)  \rbrack \nonumber \\
&  + K_{gg}^{\text{NLO}} \times \text{C} \lbrack
 \sigma_{gg} (pp\to hZ)  \rbrack ~~,
\eee
for the boosted cross section and
\bee
\widetilde{\sigma}_{\text{BDRS}}  = & K_{\overline{q}q}^{\text{NNLO}} 
\times \text{B} \lbrack \text{C} \lbrack \sigma_{\overline{q}q} (pp \to hZ, h\to \overline{b} b) \rbrack \rbrack  \nonumber \\
&  +K_{gg}^{\text{NLO}} \times \text{B} \lbrack \text{C} \lbrack \sigma_{gg} (pp \to hZ, h\to \overline{b} b) \rbrack \rbrack ,
\eee
for the cross section after applying the BDRS analysis.

Comparing the two methods we find
$\widetilde{\sigma}_{\text{Cuts}}/\sigma_{\text{Cuts}} \approx 1.09$,
constituting a $\sim 9 \%$ enhancement to the total Higgs associated
production cross section after applying a typical set of cuts for
boosted Higgs production at the LHC.\footnote{Specifically, for the
  quark-initiated contribution we have calculated the $p_T$
  distribution at LO, rather than NLO, however, due to the
  factorization of the dominant QCD correction, this has no impact on
  the comparison between gluon-initiated and quark-initiated
  distributions, which is the focus of this work.}  This arises as a
greater fraction of the gluon-initiated events survive the selection
cuts than for quark-initiated events, which can be understood from the
$p_T$ distribution in Fig.~\ref{fig:ptsm} where, for a $p_T$ cut at
$200$ GeV, a greater fraction of the total gluon-initiated events will
remain than for the quark-initiated events simply because the
gluon-initiated distribution is peaked at greater $p_T$ than the
quark-initiated distribution.

For the BDRS analysis we find
$\widetilde{\sigma}_{\text{BDRS}}/\sigma_{\text{BDRS}} \approx 0.99$
showing that the previous effect is almost completely offset because a
smaller fraction of gluon-initiated events survive the BDRS analysis
than with Drell-Yan-initiated events. This offset is, however,
dependent on the cuts and analysis applied so the effects of including
the gluon-initiated contribution must be calculated for each
independent analysis.

These numbers deserve some additional comments, since the 
interpretation of Eq.~\ref{eq:keff} is not entirely 
straightforward. The $K^{\text{eff}}$ reweighting does not include 
the different gluon acceptance, hence leads to an increased cross 
section after cuts. Once the differential acceptance is included, 
Fig.~\ref{fig:ptsmana}, this artificial enhancement becomes weaker.

The theory uncertainties on the total associated production cross
section at the LHC are $\sim 5.4 \%$ \cite{ggasshzqcd}, hence if one
applies only the boosted selection cuts this previously unconsidered
effect shifts the total cross section by almost $2 \sigma$ relative to
the assumed theory errors, however the shift is negligible if the BDRS
analysis is also applied although this is an accidental cancellation
and is not guaranteed to persist for different energies, selection
cuts, or subjet methods.  Thus to reduce theoretical uncertainty in
signal estimation at the LHC it is clearly important to include
distributions for both quark-initiated and gluon-initiated associated
Higgs production, particularly in the boosted regime.

\section{Implications for New Physics}
\label{sec:newphysics}
It is clear that gluon-initiated associated production contributes
significantly to associated production in the boosted regime.  Within
the SM this is of interest, however there are important consequences
for searches for new physics in the Higgs sector.  New physics can
potentially modify associated Higgs production at the
LHC~\cite{robert,Nishiwaki:2013cma}. The quark-initiated amplitude may
be altered at LO through modified Higgs couplings or at NLO through
the influence of new particles in loops \cite{Englert:2013tya}.
Similarly the gluon-initiated $gg \to hZ$ amplitude may also be
altered either through modified Higgs couplings to SM states, through the influence of new heavy coloured states in loops, or new s-channel pseudoscalars \cite{robert}.
Possibilities and scenarios for new states in the gluon-initiated
amplitude are multifarious and a complete study is beyond the scope of
this work hence we will only consider the case of modified Higgs
couplings in detail.\footnote{It would be interesting to calculate the
  effects of composite fermionic top partners
  as they would not only lead to additional corrections at the
  inclusive level but would also introduce new mass-thresholds into
  the $p_T$ distribution with interesting implications for different
  $p_T$ cuts.  It has been shown that loops of supersymmetric stops do not modify the gluon-initiated associated production cross section \cite{kniehl}.}

There has been a great deal of attention devoted to searching for new
physics in the Higgs sector by modifying the electroweak couplings
away from their SM values
\begin{equation} 
  g_i\to g_i (1+\delta_i) = g_i c_i
\end{equation}
either in an uncorrelated way~\cite{couplings}, or by including the
correlations present in some models such as 2HDMs~\cite{tilman}, and
fitting to the observed Higgs data.  Notwithstanding the theoretical
shortcomings of such parameter rescalings related to gauge
invariance, unitarity, and renormalisability, this procedure is often
effective in constraining the effects of UV complete models.
Ambiguities arise at NLO with Higgs coupling parameter rescalings due
to the necessity of counterterms, whose structure is intimately
related to the underlying gauge invariance of the electroweak sector.
However, these issues can be avoided if only LO processes are
considered in simple hypothesis tests to establish constraints.
Fortunately, although the $gg\to hZ$ amplitude arises at one loop, and
is technically an NNLO correction to associated production, this is a
finite LO effect and the parameter rescaling procedure can be treated
in exactly the manner as for the $gg \to h$ and $h \to \gamma \gamma$
amplitudes.

Studying Fig.~\ref{fig:feynman} it is clear that the $gg\to hZ$
amplitude is sensitive to the $hZZ$, $h \overline{b}b$, and $h
\overline{t}t$ couplings.  Also, in many UV complete scenarios with
modified $hZZ$ couplings, such as 2HDMs, gauge invariance dictates
that the $h G^0 Z$ couplings are modified by the same factor as the
$hZZ$ coupling, however the Goldstone couplings to $G^0 VV$ and $G^0
\overline{f} f$ remain as in the SM, hence we choose this as our
convention for Goldstone couplings.

Before moving on to a quantitative analysis it is worth pausing to
consider the qualitative consequences for new physics encountered when
including gluon-initiated events in boosted Higgs analyses.  To the authors
knowledge, thus far all of the many and varied studies of Higgs
couplings in new physics scenarios have assumed that all of the signal
in the boosted associated production channels arises from the
quark-initiated process
\be
\sigma(pp\to hZ) \sim \sigma (\overline{q} q \to hZ) \propto c_V^2
\label{eq:naive}
\ee
where the integration over parton distribution functions and the usual
cuts appropriate to the boosted regime are implied.  However, from
this analysis it is clear that, due to the non-negligible
gluon-initiated contribution, in reality we have
\begin{eqnarray}
  \sigma(pp\to hZ) & \sim & a_{\overline{q} q} \sigma (\overline{q} q \to hZ) + a_{gg} \sigma (gg \to hZ) \\
  &\propto& b_{\overline{q} q} c_V^2 + b_{gg} f (c_V, c_t)
\end{eqnarray}
where the $a$'s and $b$'s are constants and we have not included
dependence on $c_b$ as the bottom-loop contributions are negligible.
While this distinction may initially seem innocuous, it is
important for constraining new physics with Higgs coupling fits to
data.

Trivially one can see from Fig.~\ref{fig:feynman} that due to the
gluon-initiated contribution then, contrary to na\"{\i}ve
expectations, even if $c_V = 0$ signal will still arise in boosted
associated production channels due to the top-loop contribution.
Another interesting consequence is that the gluon-initiated
contribution is sensitive to the \emph{sign} of $c_t$ due to
interference between the triangles and boxes.  This provides an
additional handle on the sign of $c_t$ complementary to the $h \to
\gamma \gamma$ amplitude which is also sensitive to the sign.
Additionally, in many modified Higgs sectors, such as 2HDMs, $c_V \leq
1$, which is intimately related to vector-boson scattering unitarity
through sum rules \cite{Gunion:1990kf}.  Thus only assuming tree-level
processes in the boosted associated production channels, as in
Eq.~\ref{eq:naive}, unavoidably leads to the artificial restriction
$\sigma(pp\to hZ) \leq \sigma(pp\to hZ)_{\text{SM}}$ underlying any
coupling fit.  However, in many modified Higgs sectors, including
again 2HDMs, it is quite common to have $c_t \geq 1$ and, by also
including the gluon-initiated contribution, then for certain parameter
regions this also allows $\sigma(pp\to hZ) \geq \sigma(pp\to
hZ)_{\text{SM}}$, circumventing the artificially imposed restriction
$\sigma(pp\to hZ) \leq \sigma(pp\to hZ)_{\text{SM}}$.  Finally, based
on precision electroweak measurements the assumption of custodial
symmetry $c_Z=c_W=c_V$ is very robust.  Assuming only quark-initiated
associated production then leads to the assumption that the associated
production processes $pp\to hZ$ and $pp\to hW$ also obey the same
symmetry.  However, if the coupling dependence of the gluon-initiated
component is included in $pp\to hZ$ then the coupling dependence of
$pp\to hZ$ and $pp\to hW$ does not exhibit custodial symmetry as the
gluon-initiated component is absent for $pp\to hW$.

\subsection{Inclusive Associated Production}
Before turning to the case of boosted associated production, which is
relevant in searches for $h\to \overline{b} b$, we will first consider
the gluon-initiated contribution to the total associated production
cross section.  This regime is relevant in searches for $pp\to hZ$
where BDRS cuts are not applied, for example in the
ATLAS~\cite{ATLAS-CONF-2013-075} and CMS~\cite{CMS-PAS-HIG-13-017}
searches for $pp\to hV,\, h\to WW^\ast$.  From Fig.~\ref{fig:feynman}
we see that the gluon-initiated cross section must be a quadratic
polynomial in $c_V$, $c_t$, and $c_b$.  The parameter dependence of
this contribution can be determined with a hadron-level calculation
for six different parameter points.  Including the $K$-factors and
omitting the negligible dependence on $c_b$ we find for $\sqrt{s} =
14$ TeV
\begin{eqnarray}
\sigma_{gg}(pp\to hZ) & = & 136 -133 \delta_t+61 \delta_t^2  -256 \delta_t \delta_V \nonumber \\ & & +406 \delta_V+332 \delta_V^2   \text{ fb} ~~.
\end{eqnarray}
The usual quark-initiated contribution is
\be
\sigma_{\overline{q}q}(pp\to hZ) & = & 847 (1+\delta_V)^2   \text{ fb} ~~.
\ee
%

\begin{figure}[t]
  \centering
  \includegraphics[height=0.32\textwidth]{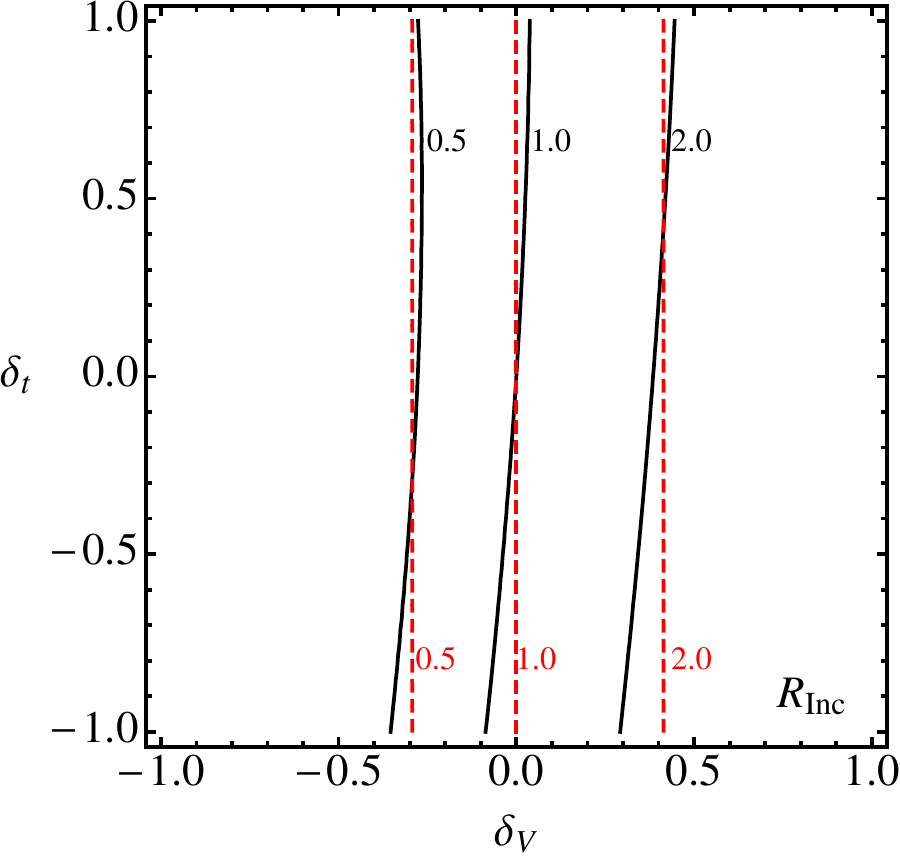}
  \caption{\label{fig:paramplotcrossinc} Parameter dependence of the
    inclusive associated production cross section relative to the SM
    result at 14 TeV after including $K$-factors and with the
    gluon-initiated contribution included (solid black) and omitted
    (dashed red).  The result which omits the gluon-initiated
    contribution is a good approximation to the full result in this
    case.}
\end{figure}

Combining both results gives the parameter dependence of the total
associated production cross section.  The SM limit agrees with the
results of \cite{ggasshzqcd} due to the $K$-factors.  Normalising the
total cross section to the SM value we have
\begin{eqnarray}
\lefteqn{R_{\text{Inc}}   \equiv   \frac{\sigma(pp\to hZ)}{\sigma(pp\to hZ)_{\text{SM}}} =} \nonumber \\  
&1  -0.14 \delta_t+0.06 \delta_t^2  -0.26 \delta_t \delta_V +2.14 \delta_V+1.20 \delta_V^2 ~~
\label{eq:ggfitinc}
\end{eqnarray}
Note that omitting the gluon-initiated component instead leads to
$R_{\text{Inc}} = (1+\delta_V)^2$ which, given the small coefficients
of the $\delta_t$ components in Eq.~\ref{eq:ggfitinc} would appear a
good approximation to the full result.  This is demonstrated in
Fig.~\ref{fig:paramplotcrossinc} where we see that, due to the small
overall contribution of the top quark loop, the dependence on
$\delta_t$ is mild, and it is reasonable to assume in this case that
the total associated production cross section is dominated by the
quark-initiated process.

\subsection{Boosted Associated Production}

Now we apply the boosted selection cuts and BDRS analysis.  We include
$K$-factors and the gluon-initiated contribution and define the
quantity $R_{\text{BDRS}}$ which contains the full associated
production cross section for $pp\to hZ, h\to \overline{b}b$ with
selection cuts and BDRS analysis applied as a function of the relevant
Higgs couplings.  The coupling dependence which is due to the
branching ratio for $h \to \overline{b} b$ is factored out in order to
make explicit the coupling dependence of the production cross section
in this channel, including the gluon-initiated process.  Specifically,
$R_{\text{BDRS}}$ is defined as\footnote{Note that in
  Eq.~\ref{eq:Rdef} the branching ratio $\text{BR}^{h\to
    \overline{b}b}$ is included in the total cross-section
  $\sigma(pp\to hZ,h\to \overline{b}b )$ hence $R_{\text{BDRS}}$ gives
  the parameter dependence of the production cross-section alone, as
  the parameter dependence of the branching ratio cancels out, by
  construction.}
\be
R_{\text{BDRS}}   \equiv   
\frac{\sigma(pp\to hZ,h\to \overline{b} b)_{\text{BDRS}}}
{\sigma(pp\to hZ,h\to \overline{b}b )_{\text{BDRS,SM}}} 
 \frac{\text{BR}^{h\to \overline{b}b}_{\text{SM}} }{\text{BR}^{h\to \overline{b}b}}
\label{eq:Rdef}
\ee
and we find
\begin{multline}
\lefteqn{R_{\text{BDRS}}   = } \\  
 1  -0.42 \delta_t+0.50 \delta_t^2  -1.41 \delta_t \delta_V +2.41 \delta_V+1.90 \delta_V^2 ~~.
\label{eq:ggfitbdrs}
\end{multline}

Comparing Eq.~\ref{eq:ggfitinc} with Eq.~\ref{eq:ggfitbdrs} we see
that in the boosted regime the dependence on $\delta_t$ is much
stronger than for the inclusive cross section.  The reason for this,
alluded to earlier, is that in the boosted regime the requirement of
larger $p_T$ essentially means that the top quark loops are probed at
CM energies close to, or in fact slightly larger than $2 m_t$.  Also,
in the high $p_T$ region the cancellation between box and triangle
diagrams is much more delicate.  Thus in the boosted regime the
contribution from the top-quark loops is enhanced relative to their
contribution in the inclusive rate.

\begin{figure}[t]
  \centering
  \includegraphics[height=0.32\textwidth]{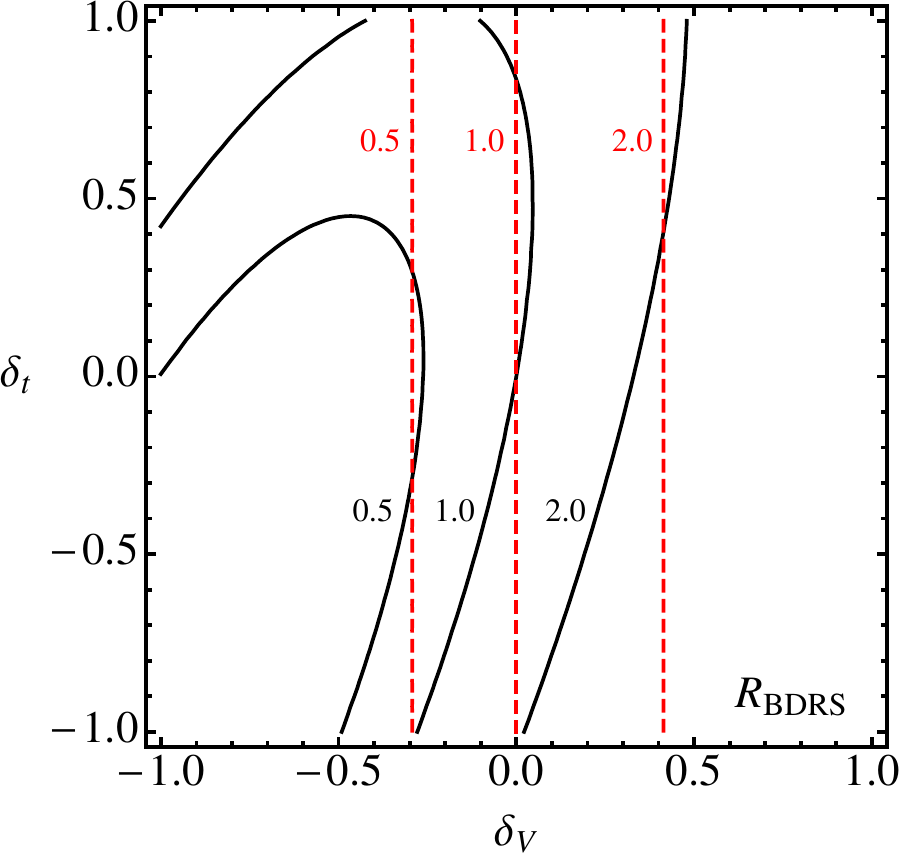}
  \caption{\label{fig:paramplotcross} Parameter dependence of
    $R_{\text{BDRS}}$ as defined in Eq.~\ref{eq:Rdef} at 14 TeV after
    including $K$-factors and BDRS cuts with the gluon-initiated
    contribution included (solid black) and omitted (dashed red).
    This is a striking example of the importance of including the
    gluon-initiated process in parameter fits involving associated
    production at the LHC.  For example, for the parameter point
    $\delta_V=0, \delta_t =-1$ ($c_V=1, c_t=0$), omitting the
    gluon-initiated process would lead to a purely SM-like
    cross-section, whereas if it is included the cross section is
    almost doubled.}
\end{figure}

It is illuminating to write Eq.~\ref{eq:ggfitbdrs} in terms of the
rescaled couplings $c_V$ and $c_t$
\be
R_{\text{BDRS}}  \approx 0.5 c_t^2 -1.4 c_t c_V +1.90 c_V^2,
\label{eq:ggfitbdrsc}
\ee
showing that at the SM point $c_V=c_t=1$ there is a mild cancellation
occurring between box and triangle diagrams.  In BSM scenarios with
modified couplings this cancellation can be disrupted, further
enhancing the role of the gluon-initiated process in boosted analyses.

In Fig.~\ref{fig:paramplotcross} we show the parameter dependence of
of $R_{\text{BDRS}}$ as defined in Eq.~\ref{eq:Rdef} at the 14~TeV LHC
with $K$-factors and BDRS cuts imposed.  As explained in the caption,
Fig.~\ref{fig:paramplotcross} clearly demonstrates the necessity of
including the gluon-initiated contribution in parameter fits involving
associated production in the boosted regime at the LHC.

While the significance of the $h \to \overline{b}b$ signals from Run~I
of the LHC is relatively weak, Run~II will lead to increased
sensitivity in this channel and understanding signals or limits on new
physics from the Run~II Higgs search data will require interpreting
the data in terms of well-motivated new physics models.  There are
many models one could consider, however we will only consider one
particularly well-motivated model with interesting modifications to
Higgs physics: the Type II 2HDM.  In addition to Higgs coupling modifications the additional heavy pseudo-scalar field present in a 2HDM may also significantly modify the gluon-initiated associated production cross section through diagrams with an additional s-channel pseudo-scalar \cite{robert}.  To demonstrate the effects of the coupling modifications in the gluon-initiated process, rather than the role of new fields, we will assume that the pseudo-scalar is decoupled and does not significantly affect rates.  This is possible in a general 2HDM for arbitrary $\alpha$ and $\beta$ parameters where there is the parameter freedom to take this limit, however the pseudo-scalar may not be taken arbitrarily heavy in the MSSM, and we reserve study of this scenario to future work.

\begin{figure}[h]
  \centering
  \includegraphics[height=0.35\textwidth]{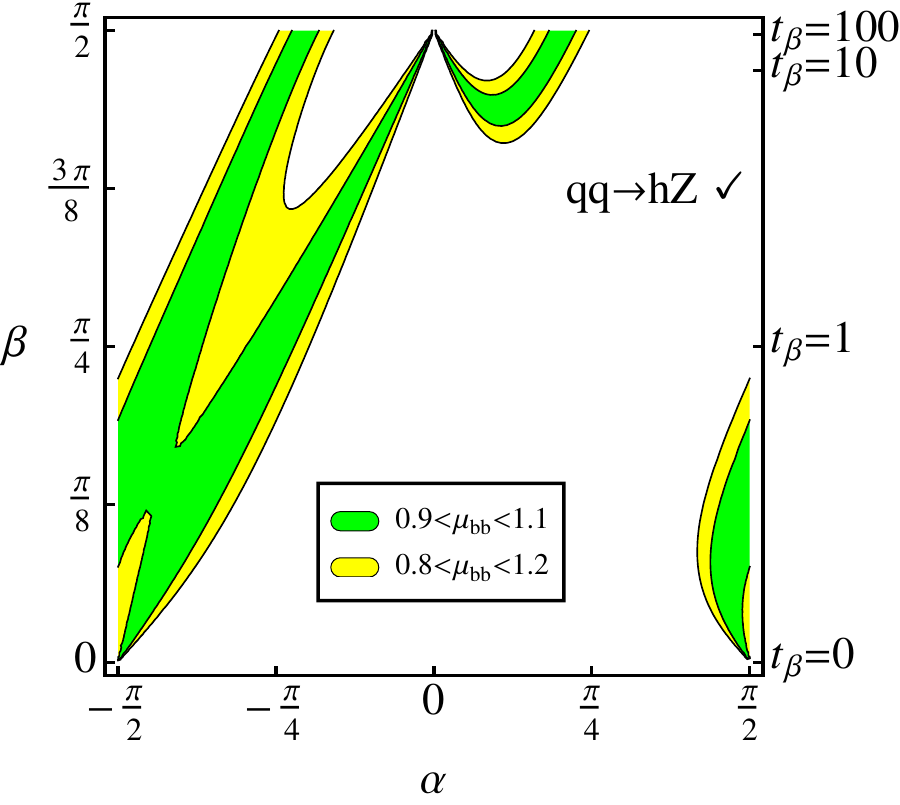}
    \includegraphics[height=0.35\textwidth]{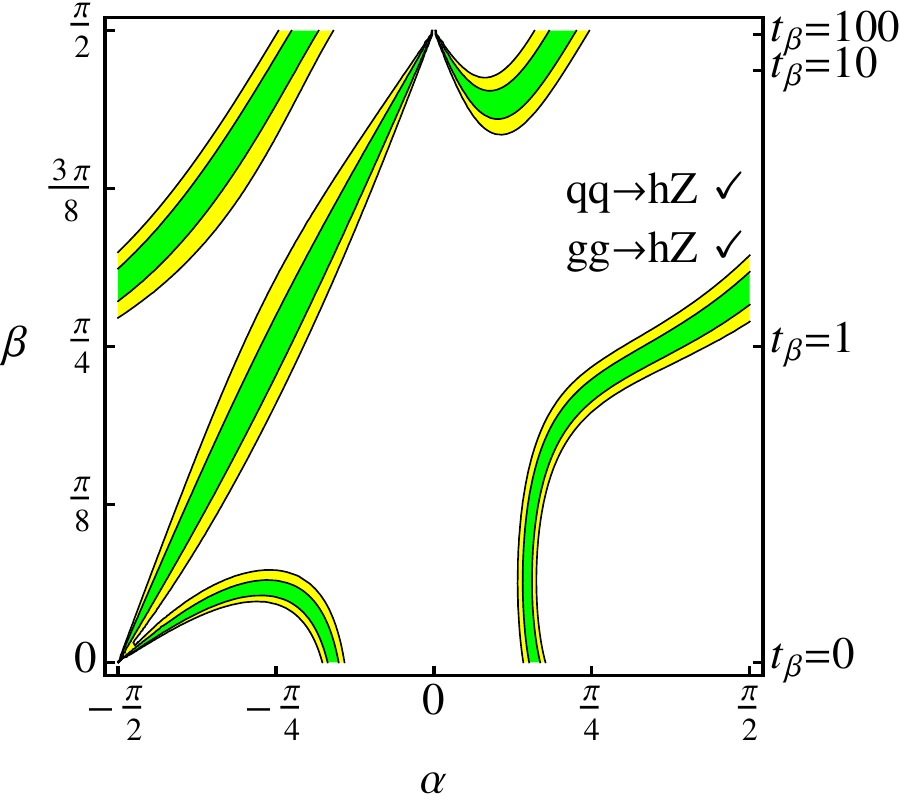}
    \caption{\label{fig:paramplot2HDM} Contours of the total signal
      strength relative to the SM in the $h\to\overline{b}b$ channel
      with BDRS analysis applied for a Type II 2HDM with the effects
      of gluon-initiated associated production omitted (above) and
      included (below).  The heavy pseudo-scalar is assumed to be decoupled, and not included in this calculation.  In this calculation we have rescaled all
      couplings, but not included triangle diagrams with $gg\to A^\ast
      \to hZ$, which may become important if the additional
      pseudoscalar $A$ is light.  Including the gluon-initiated
      associated production effects leads to significant modifications
      of the total $h\to\overline{b}b$ rate at the LHC in the type II
      2HDM.  In particular, once these effects are included,
      deviations in the total rate are more rapid as one moves away
      from the decoupling limit $\alpha = \beta-\pi/2$.  This is due
      to the rapid growth in the gluon-initiated cross-section away
      from the SM Higgs couplings as the cancellation between boxes
      and triangles is spoiled.}
\end{figure}

In Fig.~\ref{fig:paramplot2HDM} we show contours of the total
$h\to\overline{b}b$ signal strength relative to the SM, denoted
$\mu_{\overline{b}b}$, at the 14 TeV LHC without (above) and with
(below) the effects of gluon-initiated processes included.  It is
immediately apparent that away from the decoupling limit ($\alpha =
\beta-\pi/2$) the SM-like regions of parameter space are significantly
different if gluon-initiated effects are included.  Near the
decoupling limit the inclusion of gluon-initiated effects leads to a
significantly smaller region of parameter space with SM-like rates
which would lead to significantly stronger constraints on 2HDM
parameter space in the case of SM-like rate in the $h\to\overline{b}b$
channel during Run~II of the LHC.  We can study the approach to the
decoupling limit by writing $\alpha - \beta = \delta -\pi/2$ and
consider the parameter dependence of $R_{\text{BDRS}}$.  Assuming we
are close to the decoupling limit and expanding to first order in
$\delta$ we find that with the gluon-initiated process omitted \be
\mu_{\overline{b} b} (\overline{q}q\to hZ) \approx 1 - \delta (0.2
\cot \beta + 0.7 \tan \beta) ~~, \ee whereas with the gluon-initiated
process included \be \mu_{\overline{b} b} (\overline{q}q, gg\to hZ)
\approx 1 - \delta (0.6 \cot \beta + 0.7 \tan \beta) ~~, \ee and the
dependence on deviations from the decoupling limit is much stronger at
low $\tan \beta$.  This is not surprising as the gluon-initiated
associated production introduces strong dependence of the cross
section on the Higgs-top quark coupling, which in this limit is given
by $c_t \approx 1 + \delta \cot \beta$.

Looking away from the decoupling limit in Fig.~\ref{fig:paramplot2HDM}
the inclusion of gluon-initiated associated production
changes the parameter dependence of the signal rate, and hence must be
included to properly investigate or constrain these regions of
parameter space in boosted $h\to \overline{b} b$ searches.

We have only considered the impact of these new effects in one
particular example, the Type II 2HDM, however it is clear that
gluon-initiated associated production will be an important
consideration in future efforts to evaluate the viability of many
beyond the Standard Model scenarios, such as all varieties of 2HDM and
many other interesting possibilities.

\section{Summary}
\label{sec:sum}
The transition from the discovery to precision phase of Higgs boson
physics at the LHC has begun.  As it lies at the heart of hierarchy
problem, there is great potential for the Higgs boson as a future
harbinger of new physics.  Maximising this discovery potential will
require greater precision and scrutiny of theoretical predictions, a
fact which is well appreciated for SM calculations, but which is also
applicable to BSM scenarios where leading-order calculations, which
are the status quo, may fail to capture important effects.
Leading-order assumptions in BSM Higgs physics may introduce the
undesirable possibility of mis-characterising, or missing altogether,
signals of new physics: a point made clear in this work.

We have demonstrated that the $p_T$ spectrum of the gluon-initiated
contribution to associated production is fundamentally different to
the dominant quark-initiated contribution.  This is due to the
threshold behaviour of the top-loop at transverse momenta $p_T \sim
m_t$.  Although technically an NNLO contribution it is important that
the $p_T$ dependence of the gluon-initiated component is included in
the estimation of Higgs boson signals in the boosted regime, rather
than using current methods which include this contribution at the
inclusive level through an overall rescaling of the quark-initiated
distribution, obfuscating the critical differences between these two
different processes under $p_T$ cuts and boosted analysis techniques.
This is relevant for SM Higgs boson searches in boosted channels.

Looking towards BSM Higgs scenarios the gluon-initiated component may
introduce sensitivity to new coloured states through loops which would
have interesting threshold behaviour at high $p_T$ due to the mass of
new states.  This sensitivity is overlooked by making
leading-order assumptions for associated production.  Furthermore, we
have explicitly demonstrated that the gluon-initiated contribution
introduces dependence of the associated production cross section on
the Higgs-top quark coupling $c_t$, especially in the boosted regime,
and this dependence can become important away from the SM limit as a
cancellation between triangle and box diagrams is spoiled, enhancing
the effect.  This new dependence on $c_t$ has important consequences
for models where the Higgs couplings are altered as in Supersymmetry,
2HDMs, and many other scenarios.  The correct interpretation of
boosted Higgs signals at the LHC in the context of BSM scenarios,
including general Higgs coupling fits, will require treatment of the
gluon-initiated contribution to boosted associated production in
addition to the quark-initiated Drell-Yan process.

{\it{Acknowledgments.}} We thank the Institute for Particle Physics
Phenomenology and the organisers of the UK BSM Higgs 2013 workshop
where parts of this work were completed. We thank John Paul Chou,
Raffaele Tito D'Agnolo, Markus Klute, Giacinto Piacquadio, Andrea
Rizzi, and Michael Spira for valuable conversations immediately prior
to submitting this manuscript. CE is supported in parts by the IPPP
Associateship programme. MM is supported by a Simons Postdoctoral
Fellowship


\end{document}